\documentclass[useAMS,usenatbib]{mn2e}

\usepackage[dvips]{graphicx}
\usepackage[dvips]{color} 
\title[Type I Migration in Radiatively Efficient Disks]{Type I
Migration in Radiatively Efficient Disks}
\author[K. Yamada and S. Inaba]{K. Yamada$^{1}$\thanks{E-mail:
k-yamada@harbor.kobe-u.ac.jp} and S. Inaba$^{2}$\thanks{E-mail:
satoshi.inaba@waseda.jp}\\
$^{1}$Center for Planetary Science, Kobe University, Hyogo, Japan\\
$^{2}$School of International Liberal Studies, Waseda University, Tokyo, Japan}
\begin{document}

\date{Accepted ; Received ; in original form }

\pagerange{\pageref{firstpage}--\pageref{lastpage}} \pubyear{2010}

\maketitle

\label{firstpage}

\begin{abstract}
We study Type~I migration of a planet in a radiatively efficient disk using global two dimensional hydrodynamic simulations. The large positive corotation torque is exerted on a planet by an adiabatic disk at early times when the disk has the steep negative entropy gradient. The gas on the horseshoe orbit of the planet is compressed adiabatically during the change of the orbit from the slow orbit to the fast orbit, increasing its density and exerting the positive torque on the planet. The planet would migrate outward in the adiabatic disk before saturation sets in. We further study the effect of energy dissipation by radiation on Type~I migration of the planet. The corotation torque decreases when the energy dissipates effectively because the density of the gas on the horseshoe orbit does not increase by the compression compared with the gas of the adiabatic disk. The total torque is mainly determined by the negative Lindblad torque and becomes negative. The planet migrates inward toward the central star in the radiatively efficient disk. The migration velocity is dependent on the radiative efficiency and greatly reduced if the radiative cooling works inefficiently.
\end{abstract}

\begin{keywords}
hydrodynamics- radiative transfer-method: numerical-type I migration: protoplanet- planetary system: gravitational interaction
\end{keywords}

\section{Introduction}

The density waves in a protoplanetary disk are excited by the gravitational interactions with a planet \citep{b5,b6}. The tidal torque acts from the disk on a planet and generally leads to the migration of the planet. This process is known as Type~I migration and plays an important role in the formation of planets. It is suggested that Type~I migration is crucial to understand the semimajor axis distribution of extrasolar planets \citep{b9}.

There are two kinds of torques acting on a planet: one is the Lindblad torque due to spiral arms in a disk and the other is the corotation torque due to the gas which periodically exchanges angular momentum with a planet on its horseshoe orbit. The Lindblad torque occurs at the Lindblad resonances and leads to the inward migration of a planet \citep{b30}. On the other hand, the corotation torque is dependent on the physical state of a disk \citep{b29,b23} and invokes the inward or outward migration of a planet. The sum of the two torque values determines the direction and magnitude of migration of a planet.

Type~I migration was studied analytically and numerically using isothermal \citep{b27,b28,b1,b25,b3} and adiabatic disks \citep{b20,b2,b21,b17,b24}. It was shown that a planet in an isothermal disk migrates inward toward a central star in a shorter time than a lifetime of a disk \citep{b25}. The rapid inward migration suggests that planets in an isothermal disk cannot survive or stay in a far region of a disk. In an adiabatic disk with the negative gradient of the entropy, the density of the gas on the horseshoe orbit of a planet increases when the gas approaches and moves away from the planet, yielding the positive corotation torque. The positive corotation torque might become large enough to change the direction of the migration of a planet to be outward \citep{b2}.

Isothermal and adiabatic disks are idealized disks and different from real disks with dissipative processes. When a planet excites density waves in a disk by the gravitational interactions, the density waves can be altered by dissipative processes. Numerical simulations and the linear calculation were carried out, taking into account viscosity and thermal conduction as dissipative processes in disks \citep{b22,b16,b31}. With the use of the linear calculation, \citet{b19} examined the one side torque acting on a planet, including viscosity expressed by the $\alpha$ prescription. It was found that the one side torque does not change greatly unless the value of $\alpha$ is larger than 0.01, which happens to be the upper limit of $\alpha$ obtained from observations of disks \citep{b7}. This is not important in laminar disks and is not included in this study.However, viscosity does play a role in sustaining the corotation torque over longer time scales, which we do not consider in this paper. \citet{b22} included thermal diffusion in a disk as another dissipative process. They showed that the corotation torque decreases with an increase in the thermal diffusion and the total torque reduces to that in an isothermal disk. A planet migrates in a different direction, depending on the strength of the thermal diffusion.

Global simulations of Type~I migration of a planet in a radiatively inefficient disk were performed by some authors, using a grid-based code \citep{b14} and a smoothed particle hydrodynamic code \citep{b34}. They showed that a planet moves outward in an optically thick disk. However, the local calculation by \citet{b18} showed that the one side torque approaches to that in the isothermal disk with an increase in the efficiency of cooling by radiation, suggesting the inward migration of a planet in a radiatively efficient disk. In this study, we examine the total torque acting on a planet by radiatively efficient disks with various values of the opacity and reveal how the total torque depends on the efficiency of the cooling by radiation and the physical structure of a disk.

This paper is organized as follows. In the section~2, we briefly describe the basic equations and the initial conditions. The dissipative term due to radiation is included in the energy equation. In the section~3, we show the results of two-dimensional hydrodynamic simulations, taking into account the effect of the cooling due to radiation. We show the dependence of the total torque on the cooling efficiency by radiation. The direction and the magnitude of the migration of a planet change depending on the thermal structure of a disk and the cooling efficiency. Summary is given in the section~4.

\section{Basic Equations and Numerical Method}

\subsection{Basic Equations}

A planet excites density waves in a protoplanetary disk and changes the density distribution of a disk. We examine the torque acting on a planet by a radiatively efficient disk. In our study, a protoplanetary disk is assumed to be optically thin because dust particles are expected to deplete owing to the formation of planetesimals and planets \citep{b18}. We consider a planet with 5~Earth masses in an optically thin disk. A planet rotates around a solar mass star in a fixed circular orbit. The distance of the planet from the star is 15~AU and denoted by $r_{\rm{p}}$.

The problem is limited to two-dimensional flow, where all physical quantities (e.g., the surface density) depend on $r$ and $\theta$, where $r$ is the distance from the star and $\theta$ is the angle between the $x$-axis and the position vector. Governing equations are the mass conservation, the Euler equations, and the energy conservation equation with a dissipative term due to radiation by small dust particles in the disk.

We use a cylindrical coordinate where the star is located at the center of the coordinate. The mass of the planet is much smaller than that of the star and we neglect the indirect term. The basic equations read
\begin{equation}
\frac{\partial \Sigma}{\partial t}+\frac{1}{r}
\frac{\partial }{\partial r} \left(r \Sigma v_{r}\right)+\frac{1}{r}
\frac{\partial }{\partial \theta}\left(\Sigma v_{\theta}\right)=0,
\end{equation}
\begin{eqnarray}
\frac{\partial }{\partial t}\left(\Sigma v_{r}\right)+\frac{1}{r}
\frac{\partial }{\partial r} \left\{ r\left( \Sigma v_{r}^2 + p\right) \right\}
+\frac{1}{r} \frac{\partial }{\partial \theta} \left(\Sigma v_{r} v_{\theta}
\right) \nonumber \\
=\frac{\Sigma v_{\theta}^2}{r} + \frac{p}{r}
- \Sigma \frac{\partial \Phi}{\partial r},
\end{eqnarray}
\begin{eqnarray}
\frac{\partial }{\partial t}\left(\Sigma v_{\theta}\right)+\frac{1}{r}
\frac{\partial }{\partial r} \left( r \Sigma v_{r} v_{\theta}\right)
+\frac{1}{r} \frac{\partial }{\partial \theta} \left(\Sigma v_{\theta}^2 + p
\right) \nonumber \\
= -\frac{\Sigma v_{r} v_{\theta}}{r}
- \frac{\Sigma}{r}\frac{\partial \Phi}{\partial \theta},
\end{eqnarray}
\begin{eqnarray}
\frac{\partial E}{\partial t}+\frac{1}{r}
\frac{\partial }{\partial r} \left\{ r v_{r} \left( E + p \right) \right\}
+\frac{1}{r} \frac{\partial }{\partial \theta} \left\{ v_{\theta} \left(
E + p \right) \right\} \nonumber \\
= -\Sigma v_{r} \frac{\partial \Phi}{\partial r}
- \frac{\Sigma v_{\theta}}{r}\frac{\partial \Phi}{\partial \theta} - 4 \kappa
\sigma_{\rm{SB}} \Sigma \left(T^4 - T_{\rm{ini}}^4 \right),
\label{BES} 
\end{eqnarray}
where $\Sigma$ is the gas surface density, $p$ and $E$ are the vertically integrated pressure and the specific energy of the gas, respectively; $v_{r}$ and $v_{\theta}$ are the radial and tangential velocities of the gas; $\kappa$ is the Planck mean opacity, $\sigma_{\rm{SB}}$ is the Stefan-Boltzmann constant, $T$ is the temperature, $T_{\rm{ini}}$ is the unperturbed initial temperature, and $\Phi$ is the gravitational potential. We consider a less massive disk and neglect the self-gravity of the gas. The gravitational potential of the star and the planet is given by
\begin{equation}
\Phi = -\frac{GM_{\odot}}{r} -
\frac{GM_{\rm{p}}}{\sqrt{r^2+r_{\rm{p}}^2-2rr_{\rm{p}} {\rm{cos}}
\psi + \epsilon^2 H_{\rm{p}}^2}},
\label{phi_eq}
\end{equation}
where $M_{\odot}$ is the mass of the star, $M_{\rm{p}}$ is the mass of the planet, $\psi$ is the angle between the $r$-vector and the $r_{\rm{p}}$-vector, and $H_{\rm{p}}$ is the scale height of the disk at the location of the planet. The scale height is given by
\begin{equation}
H_{\rm{p}} =
\frac{\sqrt{2} c_{\rm{p}} }{ \Omega_{\rm{p}} },
\label{scaleheight_eq}
\end{equation}
where $c_{\rm{p}}$ and $\Omega_{\rm{p}}$ are, respectively, the isothermal sound velocity and the Keplerian angular velocity at $r_{\rm{p}}$. The smoothing length parameter, $\epsilon$, is introduced in order to include the effect of the scale height of the disk. \citet{b25} pointed out that the gravity of disk gas on a planet in a two-dimensional disk is greatly enhanced, compared with that in a three-dimensional disk. In this study, $\epsilon$ is set to be 0.2 for a planet with $5$~Earth masses (See Appendix).

\citet{b19} made the linear calculation and found that the one side torque exerted on a planet by a viscous disk converges to a well defined value as long as the viscosity is small. We neglect viscosity of the disk gas, even though viscosity of the gas prevents the corotation torque from saturation. We further assume that the temperature of the disk gas is the same as that of the dust particles. The dissipation term in the energy equation is derived assuming the local thermal equilibrium in an optically thin disk. The efficiency of the energy transport by radiation is mainly determined by the Planck mean opacity. When a large number of small dust particles absorb light in a disk, a disk becomes opaque. Following the formation of large particles by coalescence of the small dust particles, the opacity of a disk decreases. The evolution of the opacity strongly depends on the growth of small dust particles. Collisions of dust particles with large impact energies might result in fragmentation of the particles. A large number of small dust particles produced by the fragmentation increase the opacity of the disk. The opacity does not decrease and even increases when brittle dust particles are considered. However, \citet{b35} showed that the opacity significantly decreases in a million years if dust particles coalesce when the collision velocities of the particles are less than 10~m/s.

The specific energy of the gas is given with the pressure and the kinetic energy:
\begin{equation}
E=\frac{p}{\gamma - 1} + \frac{1}{2}\Sigma \left( v_{r}^2 + v_{\theta}^2
\right),
\label{E_eq}
\end{equation}
where $\gamma$ is the ratio of the specific heats at constant pressure and volume. \citet{b15} obtained $\gamma$ of a two-dimensional disk as $\gamma = 4/3$. The gas is assumed to be ideal and the equation of state is given by
\begin{equation}
p=\frac{\Sigma k_{{\rm{B}}} T}{ \mu m_{\rm{H}}},
\label{EOS}
\end{equation}
where $k_{{\rm{B}}}$ is the Boltzmann constant, $m_{\rm{H}}$ is the mass of a hydrogen atom, and $\mu$ is the mean molecular weight of the gas. We set up $\mu =$2.34.

The angular momentum of the disk gas is transferred to the planet. The transfer rate of the angular momentum from the disk gas at $r$ to the planet is given by
\begin{equation}
\Gamma_{r} = \int_0^{2 \pi} \Sigma
\frac{\partial  \Phi}{\partial \theta} r{\rm{d}}\theta .
\label{Gamma_den_def}
\end{equation}
By integrating the torque density over the radial distance, we obtain the total torque acting on the planet from the disk:
\begin{equation}
\Gamma = \int_{r_{\rm{min}}}^{r_{\rm{max}}} \Gamma_{r} {\rm{d}}r,
\label{Gamma_def}
\end{equation}
where $r_{\rm{max}}$ and $r_{\rm{min}}$ are, respectively, the outer and inner radii of the disk.

For the later convenience, we write all the quantities in a non-dimensional form using the unit length $r_{\rm{p}}$, the unit mass $M_{\odot}$, and the unit time $\Omega_{\rm{p}}^{-1}$. \citet{b25} made the linear calculation of the gravitational interactions between a planet and an isothermal disk. They obtained the torque on the planet exerted by the disk and found that the torque has the magnitude of $\Gamma_0 \equiv (M_{\rm{p}}/M_{\odot} )^2 (r_{\rm{p}} \Omega_{\rm{p}} / c_{\rm{p}} )^2 \Sigma_{\rm{p}} r_{\rm{p}}^4 \Omega_{\rm{p}}^2 $, where $\Sigma_{\rm{p}}$ is the surface density at $r_{\rm{p}}$. In this study, the total torque and the torque density are normalized by $\Gamma_0$ and $\Gamma_0 / r_{\rm{p}}$, respectively. The normalized quantities are denoted with a tilde (e.g., $\tilde{\Gamma}$).

\subsection{Numerical Method}

We consider a disk whose inner and outer radii are given by $r_{\rm{min}}=$6~AU and $r_{\rm{max}}=$24~AU. We use an equidistant grid in $r$ and $\theta$ directions with the resolution of $576 \times 3072$. The killing boundary conditions \citep{b4}, where all components are relaxed towards their initial state, are used in the inner zone (6~AU to 7.5~AU) and in the outer zone (21~AU and 24~AU) to avoid wave reflections from the boundaries. All the quantities in the inner and outer boundaries are always fixed to be the initial values.

Initially the temperature and the surface density distributions of the disk are given by
\begin{equation}
T_{\rm{ini}} = T_{\rm{p}} \tilde{r}^{-q}
\label{teq}
\end{equation}
and
\begin{equation}
\Sigma_{\rm{ini}}= \Sigma_{\rm{p}} \tilde{r}^{-p},
\label{sigeq}
\end{equation}
respectively, where $T_{\rm{p}}$ is the temperature at the location of the planet and set to be 72~K. The scale height of the disk, $H$, is given by
\begin{equation}
H = 9.3 \times 10^{-2} \left( \frac{T_{\rm{p}}}{72{\rm{K}}} \right)^{-1/2} \tilde{r}^{(1-q)/2} r_{\rm{p}}.
\label{scaleheight2}
\end{equation} 
The surface density, $\Sigma_{\rm{p}}$, is determined such that the total mass of the disk becomes the same as that of the standard disk model, $5\times 10^{-3} M_{\odot}$ \citep{b8}. The initial rotational velocity of the gas is calculated from the force balance in the $r$-direction: the gravity of a star, the centrifugal force, and the radial pressure gradient, while the initial radial velocity is zero. Using obtained spectral energy distributions of disks around young stellar objects, \citet{b13} showed that the power-law indices of the temperature and surface density distributions of disks range from 0.5 to 0.7 and from 0 to 1.0, respectively, that is, $0.5 < q < 0.7$ and $0 < p < 1$. We calculate the corotation torque and the Lindblad torque acting on the planet by the disk with various combinations of the two parameters, $p$ and $q$, which covers the parameter space suggested by the observations.

We develop a two dimensional global hydrodynamic program with the gravitational forces of a star and a planet and a dissipative term by radiation. The basic equations are solved simultaneously using the finite volume method with an operator splitting procedure. The source terms are computed with a second order Runge-Kutta scheme, while the advection terms are calculated with a second order MUSCL-Hancock scheme and an exact Riemann solver \citep{b26, b12}. The isothermal Riemann solver is used to calculate the total torque on a planet in an isothermal disk. It is noted that a disk with the fixed temperature distribution is hereafter called an isothermal disk in our study.

Before calculating the torque on the planet in a disk, let us introduce the cooling time, $t_{\rm{cool}}$, given by $c_{\rm{V}}/16 \sigma_{\rm{SB}} T_{\rm{ini}}^3 \kappa$, where $c_{\rm{V}}$ is the specific heat at constant volume. Furthermore, we introduce the compression time, $t_{\rm{comp}}$. A gas element is compressed while it approaches the planet from its front on the horseshoe orbit. \citet{b23} investigated the motion of gas on a horseshoe orbit of a low mass planet and obtained the width of the horseshoe region. The half width of the horseshoe region is given by
\begin{equation}
\tilde{r}_{\rm{s}} = 2 \sqrt{\left(\frac{M_{\rm{p}}}{M_{\odot}}\right) \left(\frac{r_{\rm{p}}}{H_{\rm{p}}}\right)}.
\label{halfwidthhorseshoeorbit}
\end{equation}
The velocity of a gas element on the horseshoe orbit is approximated by $1.5 \Omega_{\rm{p}} r_{\rm{s}}$. Assuming that a gas element is compressed while crossing the horseshoe region ($\sim 2r_{\rm{s}}$), we have $t_{\rm{comp}}=4/(3\Omega_{\rm{p}})$. We define the cooling efficiency by the ratio of the compression time to the cooling time:
\begin{equation}
\zeta = \frac{t_{\rm{comp}}}{ t_{\rm{cool}} }
= \frac{64 \sigma_{\rm{SB}} T_{\rm{p}}^3 \kappa}{3c_{\rm{V}} \Omega_{\rm{p}}}.
\label{zeta_def}
\end{equation}
The cooling efficiency indicates how efficient the heat is dissipated from the gas element while the gas element approaches to the planet and undergoes the compression. \citet{b18} estimated the disk opacity to be 0.0075$\rm{cm^2/g}$ when a disk has dust particles with $1$cm-radius. The cooling efficiency of the disk is given by
\begin{equation}
\zeta = 18\left( \frac{\kappa}{0.0075{\rm{cm^2/g}}} \right).
\label{zeta_eq}
\end{equation}
In an adiabatic disk, $\zeta=0$. A disk approaches an isothermal disk with an increase in $\zeta$ due to the efficient cooling of the disk.

\section{Results}

\subsection{The torque density exerted on a planet}

We study the gravitational interactions between the planet and a disk numerically. The gravity of the planet generates density waves in a disk inside and outside the orbit of the planet. The inner and outer trailing density waves pull the planet gravitationally. The positive and negative torques by the inner and outer density waves, respectively, act on the planet. In a standard disk, the gravity of the outer wave is stronger than that of the inner wave because the location of the outer density wave is closer to the planet due to the pressure gradient of a disk, leading to the negative Lindblad torque \citep{b30}. The negative Lindblad torque decreases the angular momentum of the planet.

There is another torque exerted on the planet by a gas element on the horseshoe orbit of the planet \citep{b2,b23}. When a gas element on the horseshoe orbit approaches the planet, the angular momentum is exchanged between the gas and the planet. This causes the positive or negative corotation torque on the planet, depending on the thermal structure of the disk. The corotation torque is dependent on the entropy distribution of the disk \citep{b2}. The entropy of a gas element in an adiabatic disk is conserved while it moves on the horseshoe orbit of a planet. In an adiabatic disk with the negative entropy gradient, a gas element on the horseshoe orbit is compressed when its orbit is changed from a slower orbit to a faster orbit, increasing the density of the gas element. The positive torque is exerted on the planet by the gas element. Therefore, the corotation torque is dependent on the entropy distribution of the disk. The initial entropy distribution is given by $S = p/\Sigma^{\gamma} \propto r^{\lambda}$, where $\lambda$ is $(\gamma -1)p-q$.

It was suggested that the corotation torque greatly decreases after the synodic period of the planet because the entropy of the gas in the horseshoe region of the planet tends to become uniform \citep{b2, b22, b37, b32}. This is called the saturation of the corotation torque. \citet{b22} and \citet{b37} showed that the corotation torque does not saturate in viscous disks because fresh gas is supplied into the horseshoe region from the outer region of the disk by viscosity. We also find the saturation of the corotation torque after the synodic period of the planet (about 30~rotations of the planet) because viscosity is not included in this study. If we include viscosity, we might not find the saturation of the corotation torque. Since we would like to study the effect of energy dissipation by radiation on the corotation torque in this study, we stop the simulations before the corotation torque saturates. Disks reach a tentative steady state in about ten rotations through successive gravitational interactions with a planet.

\begin{figure*}
\centerline{\includegraphics[scale=0.5,clip]{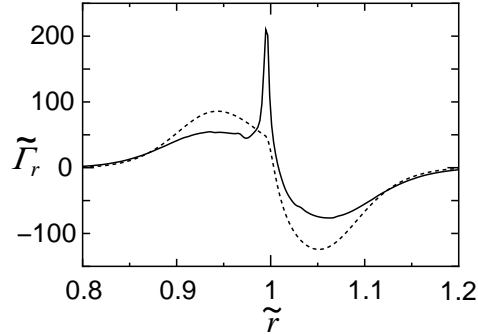}}
\caption{
The torque density acting on the planet with 5~Earth masses at $t/t_{\rm{p}} = 20$, where $t_{\rm{p}}$ is the rotational period of the planet. The disk has the density and temperature distributions with $p=0.8$ and $q=0.7$, which gives the negative power-law index ($\lambda = -0.43$) of the entropy distribution. The torque densities by the adiabatic and isothermal disks are shown by the solid and dashed curves, respectively. The positive corotation torque appears due to the negative gradient of the entropy distribution in the adiabatic disk. The torque density becomes zero in a far place from the planet due to the weak gravitational interactions. We plot the torque density in the range of $0.8 < \tilde{r} < 1.2$, even though we calculate the gravitational interactions between the planet and the gas disk with $0.4 < \tilde{r} < 1.6$.
}
\label{a8b7adiso}
\end{figure*}

We first consider the planet in the isothermal and adiabatic disks to find the difference of the torques in the idealized disks. The disks have the initial density and temperature distributions with $p=0.8$ and $q=0.7$, which gives the negative entropy gradient, $\lambda = -0.43$. Fig.~$\ref{a8b7adiso}$ shows the radial distribution of the torque density, the equation~($\ref{Gamma_den_def}$), exerted on the planet by the isothermal and adiabatic disks at $t/t_{\rm{p}} = 20$, where $t_{\rm{p}}$ is the rotational period of the planet. The solid and dashed curves correspond to the torque densities exerted on the planet by the adiabatic and isothermal disks, respectively. The large corotation torque is found in the adiabatic disk. The corotation point is located at $ \tilde{r}_{\rm{c}} = 1 - 6 \times 10^{-3}$, where $\tilde{r}_{\rm{c}}$ is the non-dimensional distance of the corotation point from the star. The torque density in the adiabatic disk becomes maximum at $\tilde{r} \simeq 1 - 5\times 10^{-3}$. The corotation torque appears in the adiabatic disk due to the enhancement of the gas density on the horseshoe orbit. Furthermore, the torque density has a local maximum and a local minimum at $ \tilde{r} \simeq 1 - 6\times 10^{-2}$ and $1 + 6\times 10^{-2}$, respectively. The distances from the planet to the local maximum and minimum are approximated by the scale height of the disk. The local maximum and minimum of the torque density correspond to the Lindblad torque. In the isothermal disk, the inner one side torque is $8.6$, while the outer one side torque is $-11.1$. As a result, the total torque exerted on the planet by the isothermal disk becomes negative ($\tilde{\Gamma} = -2.5$). The planet migrates inward toward the star in the isothermal disk. On the other hand, the large corotation torque found in the adiabatic disk becomes larger than the negative Lindblad torque and, as a result, the total torque becomes positive ($\tilde{\Gamma} = 0.3$). The planet in the adiabatic disk migrates outward.

\begin{figure*}
\centerline{\includegraphics[scale=0.5,clip]{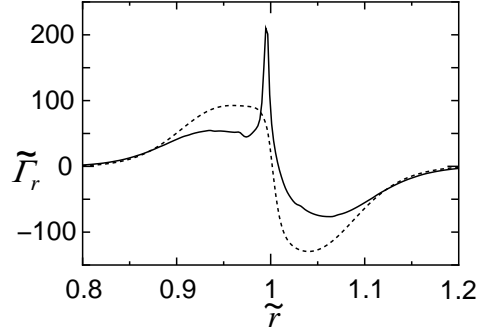}}
\caption{
The same as described in the legend to Fig.~$\ref{a8b7adiso}$, but the dashed curve is the torque density by the radiative disk with $\zeta = 18$.
}
\label{a8b7adrad1}
\end{figure*}

We further perform a simulation and calculate the torque density exerted on the planet by the disk that dissipates the energy by radiation to examine the effect of the energy dissipation on the planetary migration process. We assume that the disk contains a number of 1~cm dust particles responsible for the energy dissipation. We adopt $\zeta = 18$ as in the equation~($\ref{zeta_eq}$). Hereafter we call a disk with the energy dissipation by radiation a radiative disk. Fig.~$\ref{a8b7adrad1}$ shows the radial distributions of the torque density exerted on the planet by the adiabatic disk and the radiative disk at $t/t_{\rm{p}} = 20$. The initial disk structure is the same as that in Fig.~$\ref{a8b7adiso}$. The solid and dashed curves correspond to the torque densities in the adiabatic and radiative disks, respectively. The corotation torque in the radiative disk becomes smaller than that of the adiabatic disk because the adiabatic compression gets weaken by the energy dissipation. As shown in Fig.~$\ref{a8b7adrad1}$, the magnitude of the Lindblad torque density in the radiative disk is larger than that of the adiabatic disk. The torque density distribution of the radiative disk becomes similar to that of the isothermal disk. The total torque becomes negative ($\tilde{\Gamma} = -2.5$), leading to the inward migration of the planet.

\begin{figure*}
\centerline{\includegraphics[scale=0.5,clip]{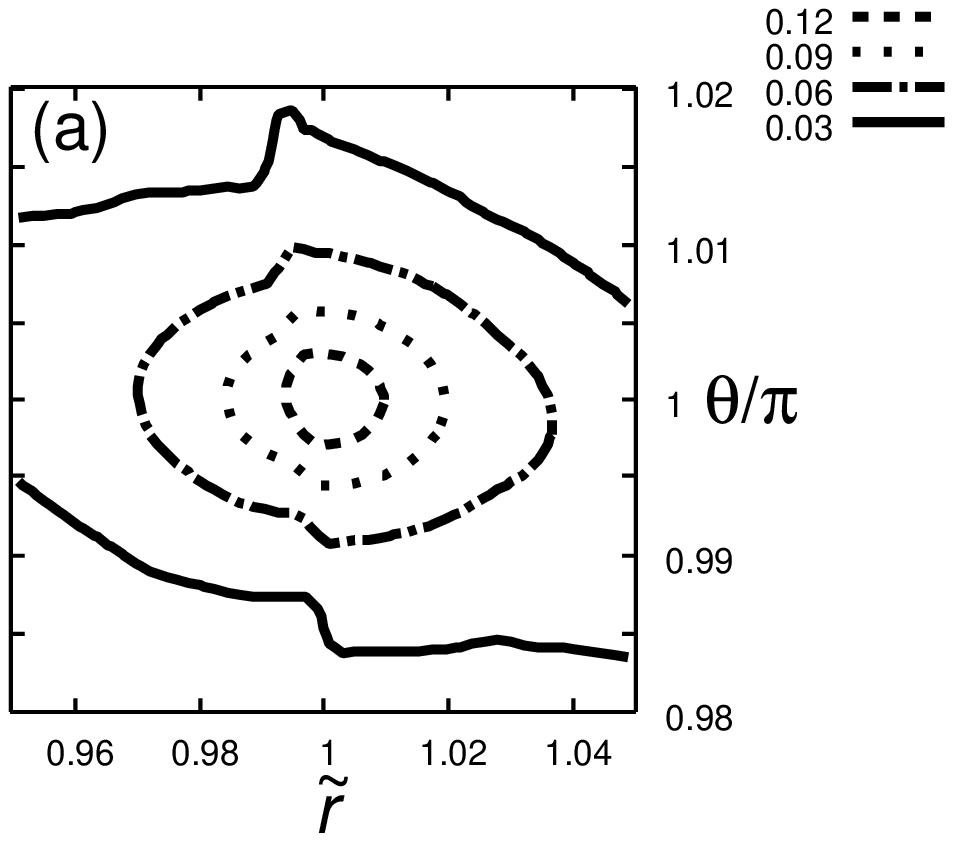}}
\centerline{\includegraphics[scale=0.5,clip]{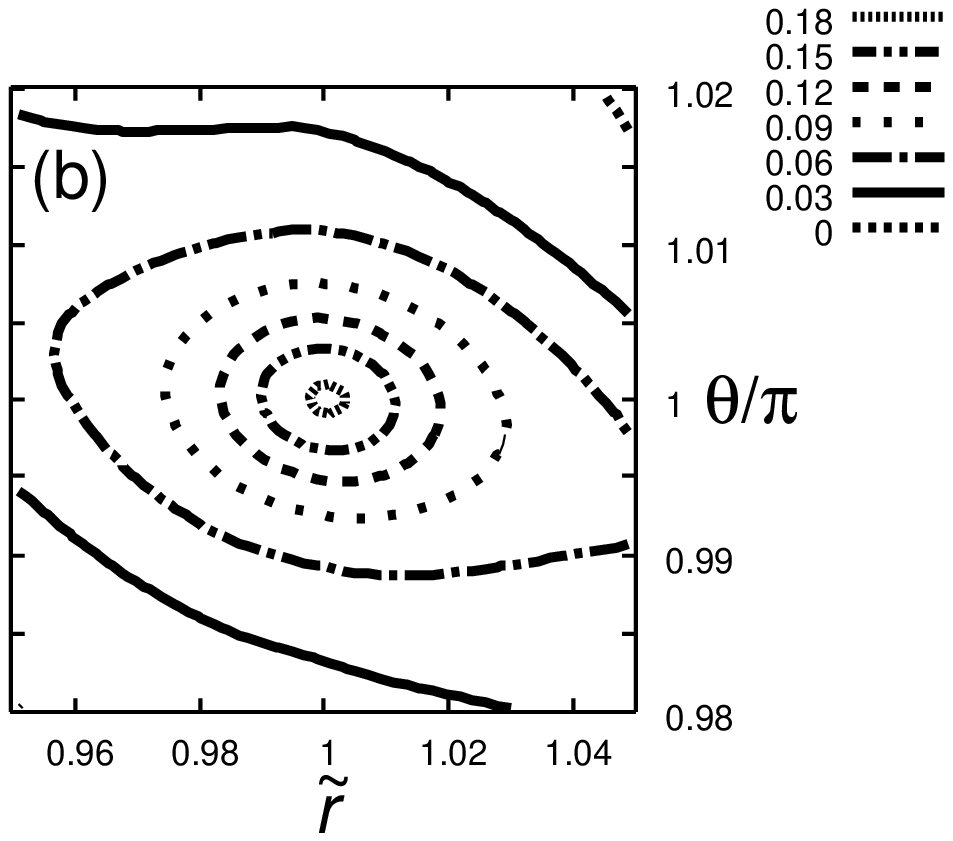}}
\caption{
The contour plot of the gas surface density around the planet at $t/t_{\rm{p}} = 20$ in (a) the adiabatic disk and (b) the radiative disk with $\zeta = 18$. The initial surface density is subtracted from the surface density for the better contrast. The densities are normalized by the initial density at $r_{\rm{p}}$. The planet is located at $\tilde{r} = 1$ and $\theta=\pi$. The gas density near the planet is enhanced in the radiative disk, compared with that in the adiabatic disk because heat is quickly radiated and the gas density is required to increase to maintain the hydrostatic equilibrium.
}
\label{den_adrad1_2d}
\end{figure*}

\citet{b18} performed the local shearing box simulations of the gravitational interactions between a planet and a radiative disk. They plotted the contour of the gas density near a planet and found the elliptical shape of the contour. The direction of the semimajor axis of the contour is inclined with respect to the direction toward a central star. This results in the larger torque density in the vicinity of a planet because the difference in the density distribution between the front and rear of the planet is larger. The contour of the density distribution in the vicinity of the planet is shown in Fig.~$\ref{den_adrad1_2d}$ at $t/t_{\rm{p}} = 20$. The panels~(a) and (b) show the density contours in the adiabatic and radiative disks, respectively. Our numerical results are qualitatively similar to that of \citet{b18}. As seen in Fig.~$\ref{den_adrad1_2d}$ (a), the gas density is enhanced in the inner horseshoe orbit region, yielding the larger positive corotation torque in the adiabatic disk. The gas density around the planet in the radiative disk becomes larger than that in the adiabatic disk. Heat generated by the compression is dissipated by radiation and the gas density is required to increase to sustain the hydrostatic equilibrium in the vicinity of the planet. Hence, the larger Lindblad torques in the radiative disk than those in the adiabatic disk are found in Fig.~$\ref{a8b7adrad1}$.

\subsection{Dependence of the total torque on the radiative efficiency and the planet mass}

\begin{figure*}
\centerline{\includegraphics[scale=0.5,clip]{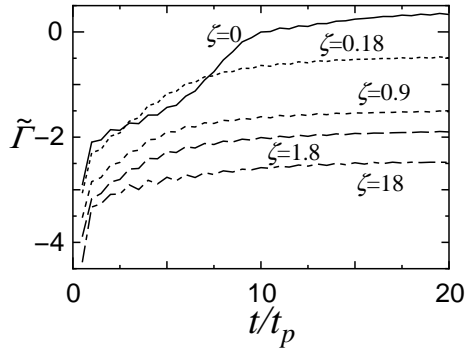}}
\caption{
The time evolution of the total torque exerted on the planet by the disks with the various radiative efficiencies: $\zeta = 0$ (solid), 0.18 (dotted), 0.9 (dashed), 1.8 (long-dashed), and 18 (dot-dashed). The disk with $\zeta = 0$ corresponds to the adiabatic disk. The total torque becomes positive in the adiabatic disk. With an increase in $\zeta$, the disk loses the energy effectively and approaches the isothermal disk. Initially all the disks have the power-law distribution of the density and the temperature with $p=0.8$ and $q=0.7$, respectively.
}
\label{a8b7evol_torq}
\end{figure*}

\begin{figure*}
\centerline{\includegraphics[scale=0.5,clip]{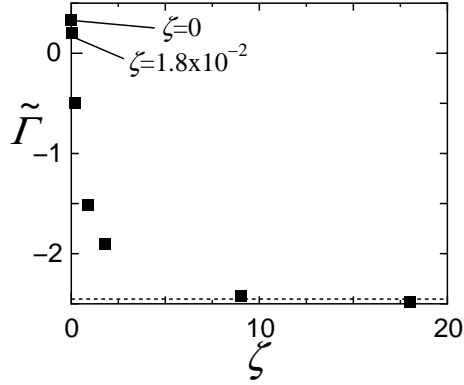}}
\caption{
The total torque exerted on the planet by the disk with the various cooling efficiencies at $t/t_{\rm{p}} = 20$. The dotted line shows the total torque exerted by the isothermal disk, which has the fixed temperature distribution with the power-law index of $q=0.7$. The total torques exerted by the disks with $\zeta>9$ is nearly the same as that by the isothermal disk.
}
\label{zeta_torq_a8b7}
\end{figure*}

We carry out a number of numerical simulations with various values of the radiative efficiency, $\zeta$. Fig.~$\ref{a8b7evol_torq}$ shows the time evolution of the total torque exerted on the planet by the disk with $p=0.8$ and $q=0.7$. The solid, dotted, dashed, long-dashed, and dot-dashed curves correspond to the total torques in the cases of $\zeta=$0, 0.18, 0.9, 1.8, and 18, respectively. The total torque increases with time at the beginning of the simulations and reaches a steady state around $t/t_{\rm{p}} = 10$. Note that the saturation of the corotation torque occurs after the synodic period ($t/t_{\rm{p}} \simeq 30$) if we continue the simulation. The total torque decreases with an increase in the radiative efficiency. The corotation torque becomes small when the heat generated by the compression is dissipated by radiation effectively. The contribution of the Lindblad torque becomes dominant and the sign of the total torque is changed from positive to negative with the increasing radiative efficiency. Except in the adiabatic disk, the total torque becomes negative and the planet moves inward toward the central star in the radiative disks. In Fig.~$\ref{zeta_torq_a8b7}$, the total torque is shown as a function of $\zeta$ for the disk with $p=0.8$ and $q=0.7$. The total torque decreases with increasing $\zeta$ and approaches that by the isothermal disk. The total torque acting on the planet by the radiative disk with $\zeta > 9$ is nearly the same with that by the isothermal disk.

\begin{figure*}
\centerline{\includegraphics[scale=0.5,clip]{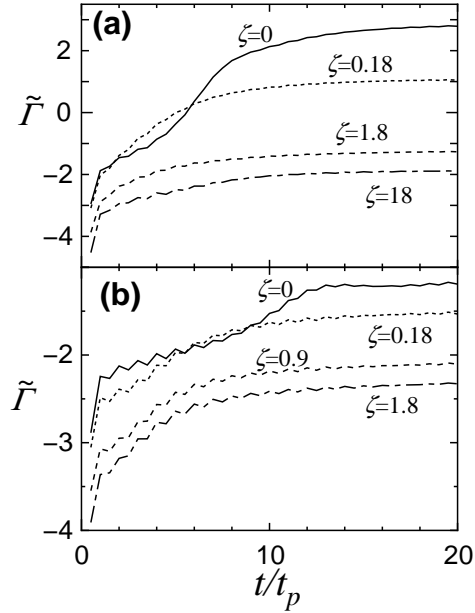}}
\caption{
The time evolutions of the total torques exerted on the planet by the two different disks with (a) $p=0.5$ and $q=1.0$ and (b) $p=1.0$ and $q=0.5$. The solid, dotted, dashed, and dot-dashed lines correspond to $\zeta=0$, 0.18, 1.8, and 18 in the panel (a) and $\zeta=0$, 0.18, 0.9, and 1.8 in the panel (b), respectively. In the panel (a), the disks with $\zeta=0$ and 0.18 exert the positive torque on the planet. In the panel (b), the total torque becomes negative even in the adiabatic disk.
}
\label{evol_torque}
\end{figure*}

Fig.~$\ref{evol_torque}$ shows the time evolutions of the total torques exerted on the planet by the radiative disks with some radiative efficiencies, $\zeta$. In the panel~(a), we use the disk with $p=0.5$ and $q=1.0$. The solid, dotted, dashed, and dot-dashed curves correspond to the total torques exerted on the planet by the disks with $\zeta=0$, 0.18, 1.8, and 18, respectively. The disk has the negative entropy gradient and the power-law index of the entropy distribution is given by $\lambda = -0.83$. This is smaller than that of the disk used in Fig.~$\ref{a8b7evol_torq}$, exerting the larger positive corotation torque on the planet. The total torques by the disks with $\zeta=0$~(the adiabatic disk) and $\zeta=0.18$ become positive due to the large positive corotation torque. The corotation torque decreases with the increasing radiative efficiency $\zeta$. The disk with $\zeta > 1.8$ exerts the negative torque on the planet. In the panel~(b), we use the disk with $p=1.0$ and $q=0.5$. The solid, dotted, dashed, and dot-dashed curves correspond to the total torques by the disks with $\zeta=0$, 0.18, 0.9, and 1.8, respectively. The power-law index of the entropy distribution is $\lambda = -0.16$, which is larger than that of the disk in Fig.~$\ref{a8b7evol_torq}$. The gas on the horseshoe orbit of the planet is compressed weakly in the disk with large $\lambda$, making the corotation torque small. The total torque becomes negative even in the adiabatic disk.

\begin{figure*}
\centerline{\includegraphics[scale=0.7,clip]{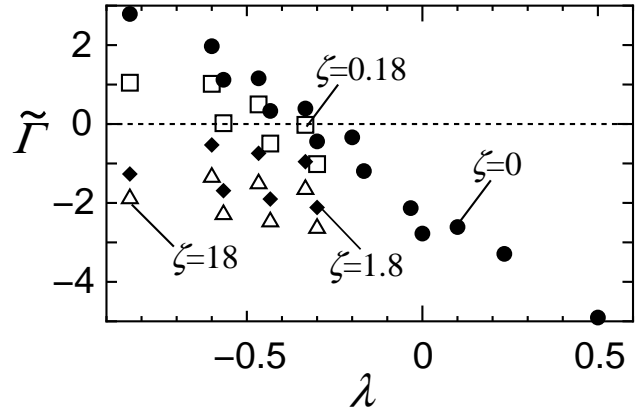}}
\caption{
The total torque exerted on the planet with 5~Earth masses by the disk as a function of the power-law index of the entropy, $\lambda$, in the cases of $\zeta = 0$ (filled circle), 0.18 (open square), 1.8 (filled diamond), and 18 (open triangle).
}
\label{total_torq_mpme5}
\end{figure*}

The total torque is dependent on the radiative efficiency, $\zeta$, and the power-law index of the entropy distribution, $\lambda$, as shown in Fig.~$\ref{total_torq_mpme5}$. We consider the disks with $p+q=1.0$ and $1.5$. The total torques at $t/t_{\rm{p}} = 20$ are plotted. The filled circles, the open squares, the filled diamonds, and the open triangles correspond to the total torques on the planet exerted by the disks with the radiative efficiencies of $\zeta=0$, 0.18, 1.8, and 18, respectively. The fluctuations of the total torques are due to the different choices of $p$ and $q$. Table~$\ref{pq_table}$ shows the list of the parameter values of $p$ and $q$ as well as the thermal state of the disk. \citet{b2} showed that the corotation torque increases with decreasing $p+q$ as well as decreasing $\lambda$. We confirm their results and find that the total torque exerted by the disks with $p+q=1.0$ is larger than that by the disks with $p+q=1.5$. The Lindblad torque is always negative in disks with the negative pressure gradient. On the other hand, the corotation torque increases to be positive with decreasing $\lambda$. The total torque becomes positive in the adiabatic disks with $\lambda < -0.4$. The planet migrates outward in the adiabatic disks with $\lambda < -0.4$.

The total torque decreases when the energy dissipation is included. The total torque on the planet by the disk with $\zeta = 0.18$ and $\lambda = -0.83$ becomes the half of that by the adiabatic disk. The energy dissipation by radiation decreases the corotation torque, resulting in the decrease in the total torque. In the disks with $\zeta = 0.18$, the sign of the total torque changes at $\lambda \sim -0.5$. When $\zeta > 1.8$, the total torque is always negative. The planet migrates inward toward the central star even in the disk, of which the amount of dust particles is one-tenth of that adopted in the standard disk. The total torque approaches that exerted by the isothermal disk with increasing $\zeta$. The total torques exerted on the planet by the disks with $\zeta=1.8$ and 18 are weakly dependent on the power-law index of the entropy distribution. The contribution of the corotation torque to the total torque becomes small. Even if the total torque by the adiabatic disk becomes positive due to the large positive corotation torque by the adiabatic compression of gas on the horseshoe orbit, it becomes negative when we take into account the energy dissipation in the disk. The planet in the disk with very small $\lambda$ and $\zeta$ might migrate outward.

\citet{b13} observed the thermal emission of dust in protoplanetary disks around T tauri stars in Taurus. They derived the disk properties of the surface density and temperature distributions. The power-law indices $p$ and $q$ range from 0 to 1 and from 0.5 to 0.7, respectively. If we only consider disks with the age over $10^6$yr, the power-law indices occupy the narrower domains of $0<p<0.6$ and $0.5<q<0.6$. Then the range of the power-law index $\lambda$ of the entropy distribution becomes $-0.6 < \lambda < -0.3$. The disks with $-0.6 < \lambda < -0.3$ exert the positive torque on the planet when the dust opacity of the disk is smaller than 1/100 of the standard dust opacity (see the equation ($\ref{zeta_eq}$)). The depletion of dust particles results in the outward or very slow inward migration of the planet in the observed disks.

Once we obtain the total torque, $\tilde{\Gamma}$, exerted on the planet by the disk, the radial migration velocity of the planet, $\dot{r}_{\rm{p}}$, is calculated as \citep{b25}
\begin{equation}
\dot{r}_{\rm{p}} = \frac{2r_{\rm{p}}\Gamma}{L_{\rm{p}}} = 0.8 \left( \frac{M_{\rm{p}}}{5 M_{\oplus}} \right) \left( \frac{\tilde{\Gamma}}{1} \right) \left[{\rm{AU/10^5yr}} \right],
\label{drp_dt}
\end{equation}
where $L_{\rm{p}}$ is the angular momentum of the planet given by $L_{\rm{p}} = M_{\rm{p}}(GM_{\odot} r_{\rm{p}})^{1/2}$ and $ M_{\oplus}$ is the Earth mass. The negative and positive radial velocities of $\dot{r}_{\rm{p}}$ correspond to the inward and outward migration of the planet, respectively. As shown in Fig.~$\ref{total_torq_mpme5}$, the total torque is on the order of $-1$ in the radiative disk with the large $\zeta$ and the planet moves about $1$~AU in $10^5$yr. The large migration velocity prohibits the planet from growing. As shown in Fig.~$\ref{total_torq_mpme5}$, the total torques in some radiative disks with small $\zeta$ happen to be on the order of $0.1$. The planet embedded in these disks migrates slowly and might be able to grow further.

\begin{figure*}
\centerline{\includegraphics[scale=0.7,clip]{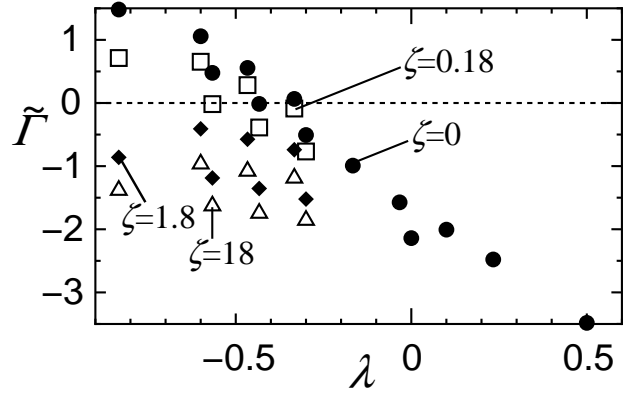}}
\caption{
The same as described in the legend to Fig.~$\ref{total_torq_mpme5}$, but the planet with 20~Earth masses.
}
\label{total_torq_mpme20}
\end{figure*}

Finally we study the dependence of the total torque on the planet mass. The total torques exerted on the planet with 20~Earth masses by the disks with various $\lambda$ and $\zeta$ at $t/t_{\rm{p}} = 10$ are shown in Fig.~$\ref{total_torq_mpme20}$. The total torque quickly reaches a steady state because the width of the horseshoe orbit of the planet is wide and it takes short time for a gas element on the horseshoe orbit to approach the planet. We find the width of the horseshoe orbit of the planet with 20~Earth masses is two times as wide as that of the planet with 5~Earth masses \citep{b36,b23}. We use the disks with $p+q=1.0$ and $1.5$. The filled circle, the open square, the filled diamond, and the open triangle correspond to the total torques by the disks with $\zeta=0$, 0.18, 1.8, and 18, respectively. The sign of the total torque on the planet with 20~Earth masses in the adiabatic disk changes around $\lambda = -0.4$, which coincides with the case of the planet with 5~Earth masses as shown in Fig.~$\ref{total_torq_mpme5}$. The zero total torques on the planets with 5 and 20 Earth masses are found at the same value of $\lambda \sim -0.6$ in the disks with $\zeta=0.18$ as well. \citet{b14} studied Type I migration of the planet with 20~Earth masses in the disk with $p=0.5$ and $q=1.6$. The disk model has the very steep entropy gradient with $\lambda = -1.4$. Due to the steep entropy gradient, the planet moves outward in the disk even with large opacity.

\begin{figure*}
\centerline{\includegraphics[scale=0.6,clip]{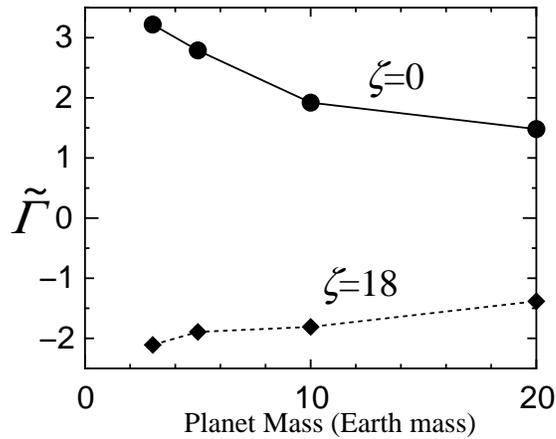}}
\caption{
The total torque as a function of the planet mass embedded in the disk with $p=0.5$ and $q=1.0$. The filled circle and diamond correspond to the cases of $\zeta=0$ and 18, respectively. The linear analysis showed that the total torque is proportional to the square of the planet mass. The total torque divided by the square of the planet mass decreases with an increase in the planet mass. This is the non-linear effect.
}
\label{planetmass_torq}
\end{figure*}

The linear analysis of the gravitational interactions between a planet and an isothermal disk showed that the total torque is proportional to the square of the planet mass \citep{b25}. Fig.~$\ref{planetmass_torq}$ shows the total torques for various planet masses in the disk with $p=0.5$ and $q=1.0$. Although the linear analysis suggested the constant magnitude, one can see from Fig.~$\ref{planetmass_torq}$ that it decreases with the increasing planet mass. This discrepancy would originate from the non-linear effect. \citet{b18} also showed the non-linear effect when large planets are considered. As compared between Figs.~$\ref{total_torq_mpme5}$ and $\ref{total_torq_mpme20}$, the total torque exerted on the planet with 20~Earth masses becomes somewhat smaller for the positive values and larger for the negative values than that on the planet with 5~Earth masses in whole.

\section{Summary}

A planet exerts the gravity on gas in a disk and changes the density distribution of the gas. Trailing density waves are generated inside and outside the orbit of the planet. The outer density wave pulls the planet gravitationally and exerts the negative torque on the planet, while the positive torque is exerted on the planet by the inner density wave. The magnitude of the gravitational force by the outer wave on the planet is usually larger than that by the inner wave due to the negative pressure gradient. The sum of the torques exerted by the outer and inner waves becomes negative, leading to the inward migration of the planet \citep{b30}. The torque exerted on the planet by the density waves is called the Lindblad torque. A gas element on the horseshoe orbit of a planet exerts torque on the planet as well. The angular momentum is exchanged between the planet and the gas element on the horseshoe orbit while the gas element moves from a slow orbit to a fast orbit. The gas element is compressed when the entropy gradient of the disk is negative. The compressed gas element increases the density and exerts the stronger gravity on the planet. The torque exerted on the planet by the gas on the horseshoe orbit is called the corotation torque. The corotation torque becomes positive, leading to the outward migration of the planet, when the entropy gradient of the disk is negative. The energy dissipation has an important role to determine the magnitude of the corotation torque. The density of a gas element on the horseshoe orbit increases the most in an adiabatic disk, in which a gas element conserves the entropy during the change of the orbit. The energy dissipation reduces the entropy of the gas element and makes the increment of the density small. The corotation torque decreases when the energy dissipation is included. We include the energy dissipation by radiation in this study. The sum of the Lindblad and the corotation torques, the total torque, determines the direction and the magnitude of the migration of the planet.

The total torques exerted on the planet by the isothermal and adiabatic disks were first examined. We find that the total torque becomes negative in the isothermal disk. The planet in the isothermal disk migrates inward toward the central star. On the other hand, the total torque becomes positive in the adiabatic disk when the entropy distribution of the disk has a steep negative slope. Gas elements on the horseshoe orbit are compressed adiabatically. The increased density during the change of the orbit from a slow orbit to a fast orbit enhances the corotation torque. The corotation torque saturates in a synodic period of the planet because viscosity is not included in the present study. The magnitude of the corotation torque is dependent on the value of the power-law index of the entropy distribution, $\lambda$. The gas disks with the negative $\lambda$ (the positive $\lambda$) exert the positive (negative) corotation torques on the planet. The total torque exerted on the planet by the adiabatic disk becomes positive when $\lambda < -0.4$ and the planet migrates outward in the disk.

We have further made a number of numerical simulations of the gravitational interactions between the planet and the disk, taking into account the energy dissipation by radiation. The density distribution of gas around the planet in the radiative disk is different from that in the adiabatic disk. More gas pressure is required to balance the gravity of the planet in the radiative disk because of the energy dissipation. The density around the planet in the radiative disk is larger than that in the adiabatic disk. The elliptical shape of an isoline of the density, whose semimajor axis is inclined with respect to the direction toward the central star, is formed near the planet. It is unclear why the isolines are more inclined in the radiative disk. Because of the inclined isoline of the density, the larger Lindblad torque is exerted on the planet in the radiative disk \citep{b18}. On the other hand, the corotation torque decreases with an increase in the energy dissipation. We consider the energy dissipation by radiation. The efficiency of the energy dissipation, $\zeta$, is defined by the ratio of the two time periods: the time period for a gas element on the horseshoe orbit to change the orbit and the time period for a gas element to lose the energy by radiation. Energy dissipates effectively in the disk with large $\zeta$. The corotation torque decreases with an increase in $\zeta$, leading to the negative total torque. The sign of the total torque is dependent on $\zeta$. The total torque invariably becomes negative in an optically thin disk with the reasonable range of $\lambda$. Even though the sign of the total torque is negative, the magnitude of the total torque might become small when the disk has small $\zeta$.

The total torque is also dependent on the mass of the planet. We examine the total torques exerted on the planets with 5 and 20~Earth masses. The total torque is approximately proportional to the square of the mass of the planet as shown by the linear analysis. With an increase in $\lambda$, the total torques exerted on the planets with 5 and 20~Earth masses by the adiabatic disk decrease and change the sign at $\lambda \simeq -0.4$. The direction of Type I migration is little sensitive to the planet mass.

\section*{Acknowledgments}
We acknowledge useful conversations with T.~Tanigawa. We also thank Nakazawa Nagare Projects members (T.~Tanigawa, A.~Nouda, Y.~Ujiie, Y.~S.~Yun, and Y.~Yamaguchi) for thoughtful comments on the simulation program. Fruitful discussions with K.~Nakazawa, H.~Emori, Y.~Nakagawa, and P.~Barge are gratefully acknowledged. We also wish to thank anonymous referee for the insightful comments to improve the paper. This study is supported by the CPS running under the auspices of the MEXT Global GCOE Program entitled "Foundation of International Center for Planetary Science." Some of the numerical simulations were carried out on the general-purpose PC farm at Center for Computational Astrophysics, CfCA, of National Astronomical Observatory of Japan.

\begin{table*}
\begin{center}
\caption{The parameter values and the thermal state of the disk used
for the simulations
\label{pq_table}}
\begin{tabular}{crrrc}
 \hline
Model number & $p$ & $q$ & $\lambda$ & Thermal State of Disk\\
\hline
1 & 1.5 & 0.0 & 0.5 & adiabatic\\
2 & 1.3 & 0.2 & 0.23 & adiabatic\\
3 & 1.2 & 0.3 & 0.1 & adiabatic\\
4 & 1.1 & 0.4 & -0.03 & adiabatic\\
5 & 1.0 & 0.5 & -0.16 & adiabatic\\
6 & 0.9 & 0.6 & -0.3 & adiabatic and radiative\\
7 & 0.8 & 0.7 & -0.43 & adiabatic and radiative\\
8 & 0.7 & 0.8 & -0.57 & adiabatic and radiative\\
9 & 0.5 & 1.0 & -0.83 & adiabatic and radiative\\
10 & 0.6 & 0.4 & -0.2 & adiabatic \\
11 & 0.5 & 0.5 & -0.33 & adiabatic and radiative\\
12 & 0.4 & 0.6 & -0.46 & adiabatic and radiative\\
13 & 0.3 & 0.7 & -0.6  & adiabatic and radiative\\
14 & 1.5 & 0.5 & 0.0   & adiabatic\\
\hline
\end{tabular}
\end{center}
\end{table*}

\appendix

\section{Deriving the softening parameter}

A planet interacts with a gas disk gravitationally. We consider a two-dimensional gas disk, in which the gas exists only in a plane. The planet in the same plane induces the larger gravity on the gas owing to the close distance. We introduce a softening parameter to incorporate the effect of the scale height of the disk. The softening parameter is used to reduce the gravity of the planet on the gas in the disk.

We assume the hydrodynamic equilibrium of the gas in the vertical direction. The velocity of the gas is independent of $z$. The temperature of the gas is assumed to be constant in the $z$-direction because an optically thin disk is considered in this study. The gravity of the planet on the three-dimensional gas is expressed by $\rho \nabla \Phi_{\rm{3D}}$, where $\Phi_{\rm{3D}}$ and the density $\rho$, respectively, are given by
\begin{equation}
\Phi_{\rm{3D}} =\frac{GM_{\rm{p}}}{\sqrt{|\textbf{r} -
\textbf{r}_{\rm{p}}|^2+z^2}}
\label{phi3d}
\end{equation}
and
\begin{equation}
\rho = \frac{\Sigma}{\sqrt{ \pi} H} \exp \left[ -\left( \frac{z}{H}
\right)^2 \right].
\label{rho3d}
\end{equation}
We integrate the gravity of the planet over $z$ as
\begin{equation}
\int_{-\infty}^{\infty} \rho \nabla \Phi_{\rm{3D}} {\rm{d}} z.
\label{int_rp_dz}
\end{equation}
Note that $\nabla$ does not include the partial derivative with respect to $z$ and $\partial / \partial z$.

The softening parameter, $\epsilon$, is defined implicitly by the following equation:
\begin{equation}
\Sigma \nabla \frac{GM_{\rm{p}}}{\sqrt{|\textbf{r} - \textbf{r}_{\rm{p}}|^2+
\epsilon^2 H_{\rm{p}}^2}} =
\int_{-\infty}^{\infty} \rho \nabla \Phi_{\rm{3D}} {\rm{d}} z.
\label{int_rp_dz}
\end{equation}
Assuming the constant scale height, $H_{{\rm{p}}}$, near the orbit of the planet, we have the following relation:
\begin{equation}
\frac{1}{\sqrt{y^2 + \epsilon^2}} = \frac{2}{\sqrt{\pi}} \int_0^{\infty}
\frac{\exp(-y^2 x^2)}{\sqrt{1+x^2}} {\rm{d}} x,
\label{soft_def}
\end{equation}
where $y$ and $x$ are defined by $|\textbf{r}-\textbf{r}_{\rm{p}} | / H_{\rm{p}}$ and $z /|\textbf{r}-\textbf{r}_{\rm{p}} |$, respectively. The softening parameter, $\epsilon$, is calculated as
\begin{equation}
\epsilon = \sqrt{ \left[ \frac{2}{\sqrt{\pi}} \int_0^{\infty}
\frac{\exp(-y^2 x^2)}{\sqrt{1+x^2}}
{\rm{d}} x \right]^{-2} - y^2}.
\end{equation}

\begin{figure*}
\centerline{\includegraphics[scale=0.7,clip]{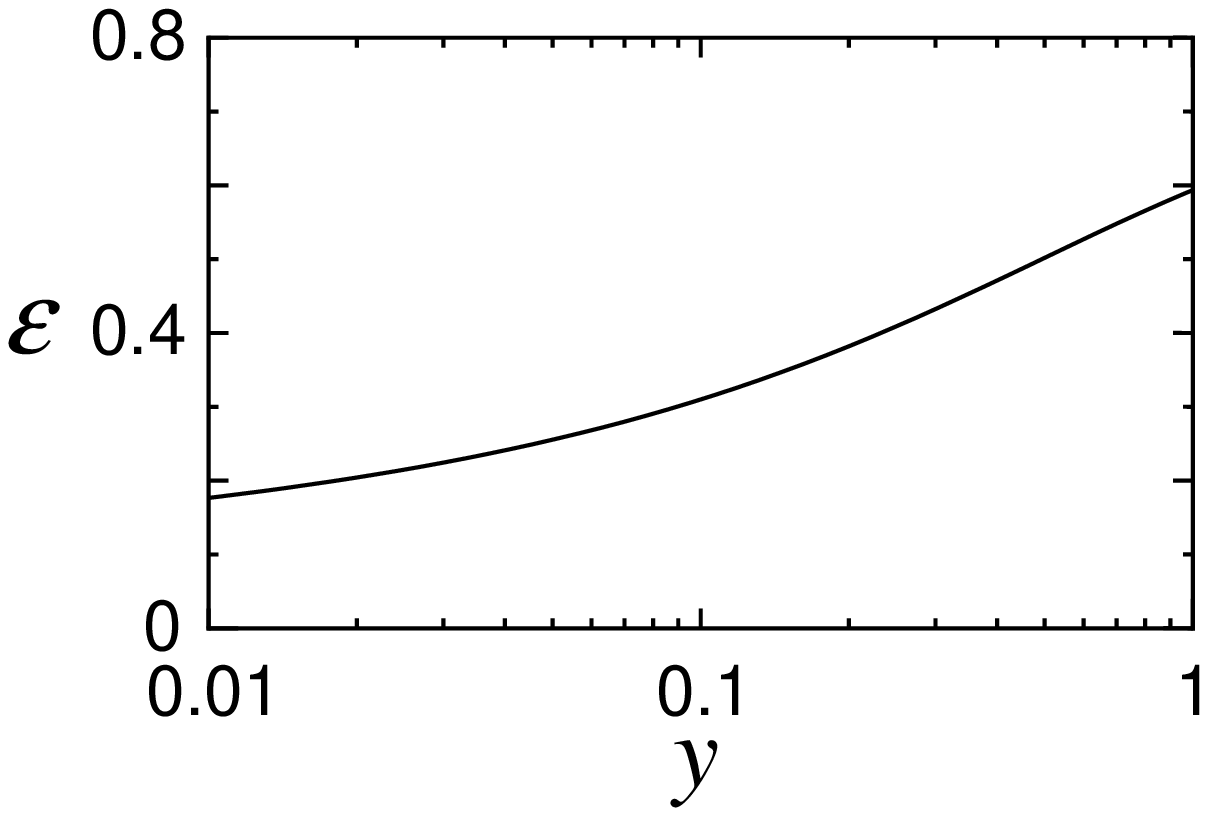}}
\caption{
The softening parameter as a function of $y$.
}
\label{sml_data}
\end{figure*}

The softening parameter is shown in Fig.~$\ref{sml_data}$. The softening parameter gradually increases with an increase in $y$. In our study, $\epsilon$ is set to be 0.2 and 0.3 for a planet with 5~Earth masses and 20~Earth masses, respectively. This softening parameter roughly corresponds to the Bondi radius ($y = 1.6\times 10^{-2}$) of the planet with 5~Earth masses and the Bondi radius ($y = 7.0\times 10^{-2}$) of the planet with 20~Earth masses.

\newpage

\label{lastpage}


\begin{thebibliography}{99}
\bibitem[\protect\citeauthoryear{Artymowicz}{1993}]{b1} Artymowicz P.,
1993, ApJ, 419, 155
\bibitem[\protect\citeauthoryear{Ayliffe \& Bate}{2010}]{b34} Ayliffe
B. A., Bate M. R., 2010, preprint, (astro-ph/10062135v1)
\bibitem[\protect\citeauthoryear{Baruteau \& Masset}{2008}]{b2}
Baruteau C., Masset F., 2008, ApJ, 672, 1054
\bibitem[\protect\citeauthoryear{Birnetiel et al.}{2009}]{b35}
Birnstiel T., Dullemond C. P., Brauer F., 2009, A \& A, 503, L5
\bibitem[\protect\citeauthoryear{Casoli \& Masset}{2009}]{b3}  Casoli
J., Masset F., 2009, ApJ, 703, 845
\bibitem[\protect\citeauthoryear{de Val-Borro et al.}{2006}]{b4} de
Val-Borro M. et al., 2006, MNRAS, 370, 529
\bibitem[\protect\citeauthoryear{Goldreich \& Tremaine}{1979}]{b5}
Goldreich P., Tremaine S., 1979, ApJ, 233, 857
\bibitem[\protect\citeauthoryear{Goldreich \& Tremaine}{1980}]{b6}
Goldreich P., Tremaine S., 1980, ApJ, 241, 425
\bibitem[\protect\citeauthoryear{Hartmann et al.}{1998}]{b7} Hartmann
L.,  Calvet N., Gullbring E., D'Alessio P., 1998, ApJ, 495, 385
\bibitem[\protect\citeauthoryear{Hayashi}{1981}]{b8} Hayashi C, 1981,
PThPS, 70, 35
\bibitem[\protect\citeauthoryear{Ida \& Lin}{2008}]{b9} Ida S., Lin
D.N.C., 2008, ApJ, 673, 487
\bibitem[\protect\citeauthoryear{Inaba et al.}{2005}]{b12} Inaba S.,
Barge P., Daniel E., Guillard H., 2005, A\&A, 431, 365
\bibitem[\protect\citeauthoryear{Kitamura et al.}{2002}]{b13} Kitamura
Y., Momose M., Yokogawa S., Kawabe R., Tamura M., Ida S., 2002, ApJ,
581, 357
\bibitem[\protect\citeauthoryear{Kley \& Crida}{2008}]{b14} Kley W.,
Crida A., 2008, A\&A, 487L, 9
\bibitem[\protect\citeauthoryear{Kley et al.}{2009}]{b32} Kley W.,
Bitsch B., Klahr H., 2009, A\&A, 506, 971
\bibitem[\protect\citeauthoryear{Li et al.}{2000}]{b15} Li H., Finn
J.M., Lovelace R.V.E., Colgate S.A., 2000, ApJ, 533, 1023
\bibitem[\protect\citeauthoryear{Li et al.}{2009}]{b16} Li H., Lubow
S.H., Li S., Lin D.N.C., 2009, ApJ, 690, 52
\bibitem[\protect\citeauthoryear{Masset et al.}{2006}]{b36} Masset F. S.,
D'Angelo G., Kley W., 2006, ApJ, 652, 730
\bibitem[\protect\citeauthoryear{Masset \& Casoli}{2009}]{b17} Masset
F., Casoli J., 2009, ApJ, 703, 857
\bibitem[\protect\citeauthoryear{Morohoshi \& Tanaka}{2003}]{b18}
Morohoshi K., Tanaka H., 2003, MNRAS, 346, 915
\bibitem[\protect\citeauthoryear{Muto \& Inutsuka}{2009}]{b19} Muto
T., Inutsuka S., 2009, ApJ, 701, 18
\bibitem[\protect\citeauthoryear{Paardekooper \& Mellema}{2006a}]{b33}
Paardekooper S.-J., Mellema, G., 2006a, A\&A, 453, 1129
\bibitem[\protect\citeauthoryear{Paardekooper \& Mellema}{2006b}]{b20}
Paardekooper S.-J., Mellema, G., 2006b, A\&A, 459, 17
\bibitem[\protect\citeauthoryear{Paardekooper \& Mellema}{2008}]{b21}
Paardekooper S.-J., Mellema, G., 2008, A\&A, 478, 245
\bibitem[\protect\citeauthoryear{Paardekooper \&
Papaloizou}{2008}]{b22} Paardekooper S.-J., Papaloizou J.C.B., 2008,
A\&A, 485, 877
\bibitem[\protect\citeauthoryear{Paardekooper \&
Papaloizou}{2009a}]{b37} Paardekooper S.-J., Papaloizou J.C.B., 2009,
MNRAS, 394, 2283
\bibitem[\protect\citeauthoryear{Paardekooper \&
Papaloizou}{2009b}]{b23} Paardekooper S.-J., Papaloizou J.C.B., 2009,
MNRAS, 394, 2297
\bibitem[\protect\citeauthoryear{Paardekooper et al.}{2010}]{b24}
Paardekooper S.-J., Baruteau C., Crida A., Kley W., 2010, MNRAS, 401,
1950
\bibitem[\protect\citeauthoryear{Tanaka et al.}{2002}]{b25} Tanaka H,
Takeuchi T, Ward W., 2002, ApJ, 565, 1257
\bibitem[\protect\citeauthoryear{Toro}{1999}]{b26} Toro E.F., 1999, in
Riemann solvers and numerical methods for fluid dynamics. A practical
introduction (Springer)
\bibitem[\protect\citeauthoryear{Ward}{1986}]{b27} Ward W., 1986,
Icarus, 67, 164
\bibitem[\protect\citeauthoryear{Ward}{1989}]{b28} Ward W., 1989, ApJ, 336, 526
\bibitem[\protect\citeauthoryear{Ward}{1991}]{b29} Ward W., 1991, LPI, 22, 1463
\bibitem[\protect\citeauthoryear{Ward}{1997}]{b30} Ward W., 1997,
Icarus, 126, 261
\bibitem[\protect\citeauthoryear{Yu et al.}{2010}]{b31} Yu C., Li H.,
Li S., Lubow S.H., Lin D.N.C., 2010, ApJ, 712, 198


\end{thebibliography}
\end{document}